%% file: main.tex
\documentclass[runningheads]{llncs}
\usepackage[T1]{fontenc}
\usepackage{graphicx}
\usepackage{algorithm}
\usepackage{graphicx}
\usepackage{algpseudocode}
\usepackage{amsmath}
\usepackage{amsfonts}
\usepackage[misc]{ifsym}
\usepackage{xspace}

\usepackage{multirow, makecell} 
\usepackage{tabularx}

\graphicspath{{figure/}}

\newcommand{\fedrec}{\textsf{FedRec}}
\newcommand{\sys}{\textsf{DP-FedRec}}

\def\ie{\textit{i.e.}\xspace}

\begin{document}

\title{A Privacy-Preserving Subgraph-Level Federated Graph Neural Network via Differential Privacy}

\titlerunning{Federated Graph Neural Network via Differential Privacy}

\author{Yeqing Qiu\inst{1,2} \and
Chenyu Huang\inst{1} \and
Jianzong Wang \inst{1}\thanks{Corresponding author: Jianzong Wang, jzwang@188.com} \and
Zhangcheng Huang\inst{1}\and
Jing Xiao\inst{1}}

\authorrunning{Y. Qiu et al.}
%
\institute{Ping An Technology (Shenzhen) Co., Ltd., Shenzhen, China \\
\and
Beijing Jiaotong University, Beijing, China \\
\email{yeqing@bjtu.edu.cn, hcyray@gmail.com, jzwang@188.com, hzcsimon@vip.qq.com, xiaojing661@pingan.com.cn}
}

\maketitle
\input{abstract}

\input{introduction}

\input{background}
\input{design}
\input{analysis}
\input{evaluation}

\input{related_work} 
\input{conclusion}

\bibliographystyle{splncs04}
\bibliography{reference.bib}

\end{document}

%% file: abstract.tex
\begin{abstract}
Currently, the federated graph neural network (GNN) has attracted a lot of attention due to its wide applications in reality without violating the privacy regulations.
Among all the privacy-preserving technologies, the differential privacy (DP) is the most promising one due to its effectiveness and light computational overhead.
However, the DP-based federated GNN has not been well investigated, especially in the sub-graph-level setting, such as the scenario of recommendation system.
The biggest challenge is how to guarantee the privacy and solve the non independent and identically distributed (non-IID) data in federated GNN simultaneously. 
In this paper, we propose \sys, a DP-based federated GNN to fill the gap. 
Private Set Intersection (PSI) is leveraged to extend the local graph for each client, and thus solve the non-IID problem.
Most importantly, DP is applied not only on the weights but also on the edges of the intersection graph from PSI to fully protect the privacy of clients.
The evaluation demonstrates \sys\ achieves better performance with the graph extension and DP only introduces little computations overhead.
\keywords{Recommendation System, Federated Learning, Subgraph-Level Federated Learning, Graph Neural Network, Differential Privacy}
\end{abstract}

%% file: introduction.tex
\section{Introduction}
Graph neural network (GNN) has been applied to multiple scenarios such as molecule prediction~\cite{fout2017protein,2013Security}, social network analysis~\cite{chen2018fastgcn,2017Privacy}, recommendation systems~\cite{jin2020multi} and knowledge graph~\cite{schlichtkrull2018modeling}.
However, GNN approaches mainly rely on the centralized data, which is different from the real-world scenario where the source data may be stored at different organizations.
For example, e-commerce platforms that sell different types of items have separate purchase and rating records of their users and items. In order to explore potential new users and provide better recommendation services to existing users, E-commerce platforms would build a better model jointly learned from multiple data resources. In the meantime, the user privacy should be protected for ethical concerns and compliance with government regulations.

As a result, approaches are presented to combine the well-known privacy-preserving framework, federated learning (FL), and GNN. 
Different technologies such as differential privacy (DP)~\cite{wu2021linkteller,wu2021fedgnn,zhou2020vertically,2020Selective}, homomorphic encryption~\cite{wu2021fedgnn}, secret sharing~\cite{zhou2020vertically} are widely applied to dealing with risks of privacy leakage.
Among the techniques mentioned above, DP is the most promising one due to its light computational overhead and high fidelity.
DP perturbs the data with a small noise without lowering the accuracy of the entire model, \ie, if the input signal changes, the distribution of the output only changes a little.


Currently, real-world scenarios of privacy-preserve graph learning mainly concentrates on three settings~\cite{he2021fedgraphnn}: graph-level setting~\cite{zhou2020vertically}, sub-graph-level setting~\cite{qi2020privacy,wu2021linkteller,yang2020secure,wu2021fedgnn} and node-level setting~\cite{cheungfedsgc}. 
Among these settings, sub-graph-level is the most attractive since it is a good fit to the most important/common application scenario such as recommendation system and knowledge graph.
For example, in recommendation systems, every data holder will only own the part of graph that contains the relationship between user and item. 
The biggest challenge in this setting is preserving the privacy and solving the Non-IID problem in federated GNN simultaneously.
However, these work either assume one party owns the global topology~\cite{he2021fedgraphnn,zhou2020vertically}, which violate the basic assumption in general scenario where no one is allowed to own the whole typology, or do not consider the information from the neighbors~\cite{wu2021fedgnn,qi2020privacy},
which do not solve the Non-IID problem and thus lead to low accuracy of the model.
Therefore, these approaches cannot be directly applied in the general sub-graph level scenario.

In this paper, we propose a novel DP-based GNN that aims at the sub-graph level setting in chapter 3.
To solve the Non-IID problem, the \fedrec\ that utilizes the K-hop extension to expand the sub-graph of each client is introduced. The privacy of the communication between clients is preserved via the Private Set Intersection (PSI).
Furthermore, we propose \sys\ that leverages DP in \fedrec. The core idea is to apply well-designed noises to both adjacency edges and weights of client's sub-graph.
Specifically, the Laplacian noise is applied on the edges via Lapgraph algorithm and apply the Laplacian noise on the weights. The analysis and evaluation in chapter 4 and 5 show the K-hop extension achieves better performance than previous schemes and the DP introduces limited computational overhead. We summarize the main contributions as follows:


\begin{itemize}
\item We propose a state-of-art learning paradigm on sub-graph setting based on DP, which is able to be applied to many link prediction tasks.  
\item We utilize K-hop extension for exchanged feature and adjacency information and preserve the privacy of both the feature and edge information via DP.
\item We evaluate our algorithms on two recommendation datasets, and demonstrate the effectiveness of our approach.
\end{itemize}

%% file: background.tex
\section{Preliminaries}
\subsection{Problem Formulation}
In this work, denote $\mathcal{U}=\{u_1,u_2,\cdots,u_n\}$ and $\mathcal{P}=\{p_1,p_2,\cdots,p_m\}$ as user and item respectively. 
The purchasing interaction of user and item relationship is represented by a bipartite graph $\mathcal{G}\in \mathbb{R}^{n\times m}$, in which the value of edges refers to the points the user rate the item. 
Since each client will only have a part of global graph, for client $i$, the user-item bipartite graph is denoted as $G^i = (V^i, E^i)$. 
In detail, the set of vertexes and edges are denoted as $V^i, E^i$ respectively. 
The task is to predict the ratings of users and items based on user-item graph. Thus, client $i$ will train a local model in round $r$, the parameters of which are denoted as $\theta_i^l$. The global model parameter that aggregate from each client is $\theta^r$.
Additionally, define $\mathsf{dist}(x, y, \mathcal{G})$ as the shortest path of vertex $x$ and $y$ in graph $\mathcal{G}$. Define $\mathsf{dist}(v, \mathcal{S}, \mathcal{G}) = \min_{x\in \mathcal{S}}{\mathsf{dist}(v, x, \mathcal{G})}$.
The notation is summarized in Table~\ref{tab:notation}.
\vspace{-5mm}
\begin{table}[htbp]
    \caption{Notations used in \sys.}
    \begin{center}
        \begin{tabular}{|c|c|}
            \hline $l$ & number of clients \\
            \hline $n,m$ & number of users and items in graph $\mathcal{G}$\\
            \hline $u_i$, $p_i$ & user $i$, item $i$ \\
            \hline $G^{i}$ & user-item graph of client $i$\\
            \hline $V^i, E^i$ & vertex set and edge set in $G^i$ \\
            \hline $\bar{G}^{i}$ & extended user-item graph of client $i$\\
            \hline $\bar{V}^i, \bar{E}^i$ & extended vertex set and edge set in $\bar{G}^i$ \\
            \hline $K$ & parameter of K-hop extension\\
            \hline $r$ & communication round\\
            \hline $\theta^r$ & parameters of global model in round $r$\\
            \hline $\theta^r_{i}$ & parameters of client $i$'s local model in round $r$\\
            \hline $\nabla\theta_{i}$ & gradient of parameters of local model\\
            \hline
        \end{tabular}
    \label{tab:notation}
    \end{center}
\vspace{-25pt}
\end{table}

\vspace{-5mm}

\subsection{Local Differential Privacy}
Local differential privacy guarantees the privacy of the user in the process of collecting information. 
Specifically, before the user uploads the data to an untrusted third party, a certain amount of noises is added to the uploaded data. This guarantees that the data collectors can hardly infer the specific information of any user, but are able to learn the statistical properties of the data by increasing the amount of data.

Different from the previous unweighted graph~\cite{wu2021linkteller}, the user-item graph in recommendation system is a weighted graph.
Therefore, in order to protect information of user-item graph, the definition of DP in undirected weighted graph data is obtained by combining both unweighted undirected graph information and weight information. Consistent with prior work~\cite{wu2021linkteller}, we adapt the idea of edge differential privacy based on adjacency matrix.
\begin{definition}[Neighbor Relation]
 Two matrices are called neighbors if there is only one different node.
 Specifically, the graph corresponding to the two matrices can be obtained by adding or deleting an edge or modifying value of an edge.
\end{definition}

\begin{definition}[$\epsilon$-Weighted Edge Local Differential Privacy]
A mechanism $M$ is called to satisfy $\epsilon$-Weighted Edge Local Differential Privacy if for all neighbor matrix pairs $X$ and $X'$, and for any possible output $t \in Range(M)$:
\begin{equation} \label{(1)}
    P[M(X)=t] \le e^\epsilon P[M(X')=t]
\end{equation}
\end{definition}




\subsection{Federated Graph Neural Network}

Graph neural network is widely used in recent recommendation systems~\cite{wu2020comprehensive}. In this paper, we leverage the graph convectional network (GCN)~\cite{kipf2016semi} under the message passing neural network framework (MPNN) .
MPNN is a supervised learning framework which extracts information from the user-item graph by aggregating adjacency information into the latent space, and then generates the prediction from the latent space.

Furthermore, we extend GNN to the federated scenario which is the same as in~\cite{he2021fedgraphnn}.
Specifically, it is a sub-graph setting where each entity/company has a part of data/graph, such as users and rating information, and a model is jointly trained on the entire data for better prediction accuracy.
Therefore, there are multiple clients and one centralized server. 
For communication round $r$ in the training stage, 
client $i$ will get the model parameter $\theta^r_i$ by training the local model for $e$ epochs on the sub-graph $G^i_r$. The server will aggregates the parameters $\theta_r=\frac{\sum\theta^r_i}{l}$ and distribute them to all local clients, and each client updates its local model parameters as $\theta_i^r$. 






\subsection{Private Set Intersection}
Private Set Intersection (PSI) is a cryptographic protocol in multiparty computation. It allows two clients to get the intersection set of the data without revealing any information outside the intersected data.
There are many different implementations of PSI, \sys\ instantiates the PSI the same as~\cite{kolesnikov2017practical}. It leverages the programmable pseudo random function (OPPRF) which is fast and efficient. 

%% file: design.tex
\section{Approach}
\subsection{Overview}
The basic federated GNN framework does not use the graph information from others and could cause non-IID problem in the training data. We will first present \fedrec\ which extends the sub-graph of each client without leaking the information of the edges. 
Then we will introduce \sys\ that combines \fedrec\ and DP and jointly considers the privacy of both weights and connectivity of the edges simultaneously.

\begin{figure}[htbp] 
	\centering
	\includegraphics[width=0.9\textwidth]{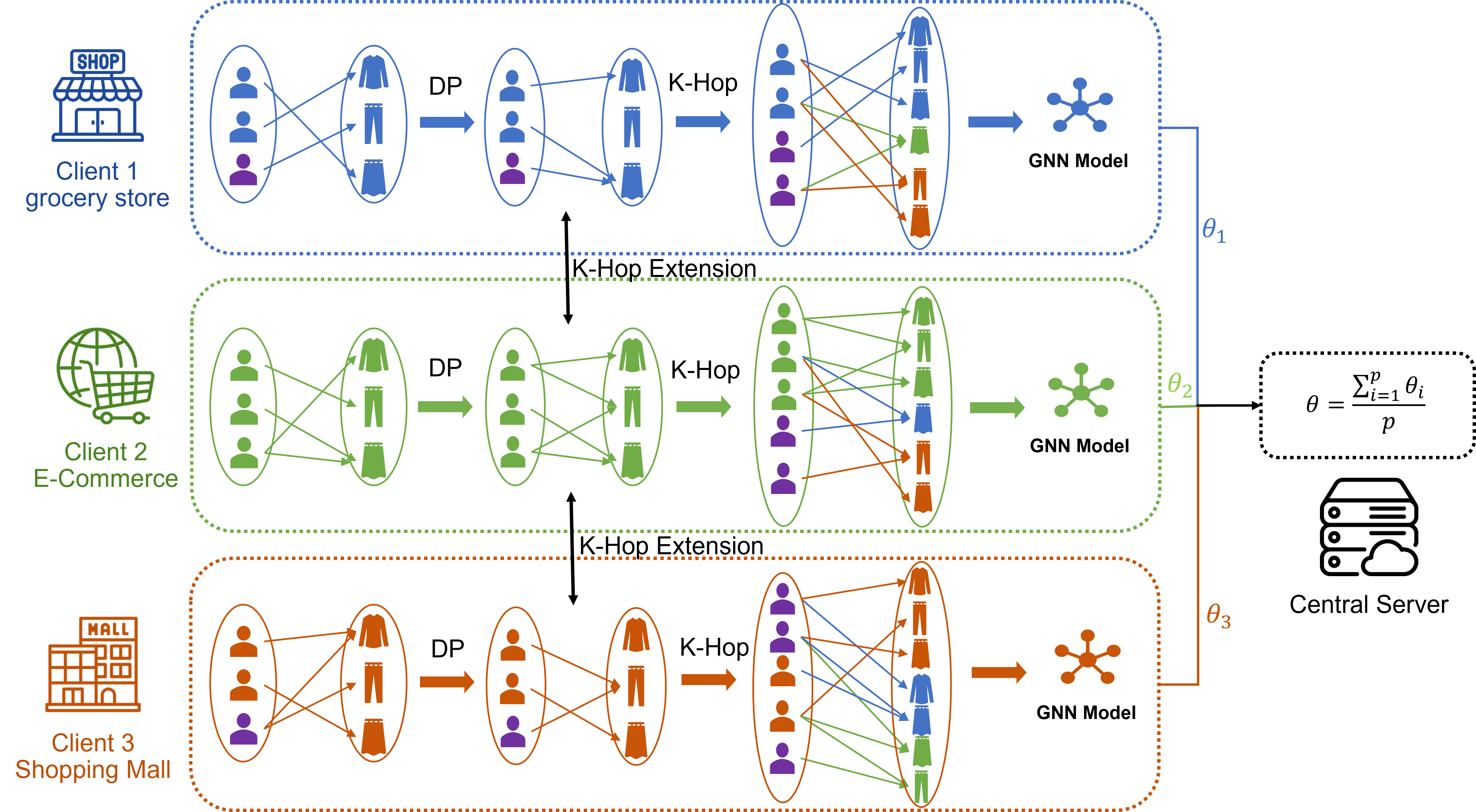}
	\caption{Overall framework of \sys. Each bipartite graph refers to purchasing relationship between user and item in each platform and client. The purple ones are the users in the intersection set of clients' sub-graphs.}
    \label{fig_fedrec} 
\end{figure}

Specifically, \sys\ jointly trains a model via four steps as shown in Fig.~\ref{fig_fedrec}: (i) All the clients add noises on the graph data including both weights and edges; (ii) The clients extend the local graph via K-hop extension; (iii) Each client trains the local model on the extended graph and submits the parameters to the server; (iv) The centralized server aggregates the parameters and distributes the updated parameters to all the clients. The process will continue until certain number of rounds is reached.




\vspace{-25pt}

\subsection{User-Item Graph K-hop Expansion}
To overcome the non-IID problem, \fedrec\ privately exchanges the edges information between clients. 
The main idea is to expand the edges from the intersected users in different sub-graphs. In two-client setting, for example, the intersected users are the users appear in both sub-graphs.
We integrate PSI to the K-hop extension, which avoid leaking the user-item information that is not in the intersection set.

Without loss of generality, suppose there are two clients, client $i$ and client $j$, who exchange the edges information via K-hop extension and generate the extended sub-graph $\bar{G}^i$ and $\bar{G}^j$.
The vertex set and edge set of user-item graph $G^i$ are $V^{i}$ and $E^{i}$ respectively, and client $j$ also records the $G^j=(V^j, E^j)$.
Firstly , client $i$ and client $j$ will execute PSI protocol to get the intersection of $V^i$ and $V^j$, denoted as $V^{i,j}$, \ie, $V^{i,j}=\mathsf{PSI}(V^i, V^j)$ (Line 4 in Algorithm~\ref{alg:k-hop}).
Then, the K-hop extension is performed by extending the edges and vertexes on $V^{i,j}$. 
The extended vertex set $\bar{V}^{i,j}$ will cover the vertexes in  $V^j$ within $K$ hops from $V^{i, j}$ (Line 5 in Algorithm~\ref{alg:k-hop}).
Similarly, the extended edge set $\bar{V}^{i,j}$ will cover the edges that both vertexes are in $\bar{V}^{i,j}$ (Line 6 in Algorithm~\ref{alg:k-hop}).
Next, the client $i$ and client $j$ will exchange the extended vertex set and extended edge set~(Line 7,10 in Algorithm~\ref{alg:k-hop}). 
Lastly, the client $i$ combines the extended vertexes and edges with $G^{i}$ to form the new graph $\bar{G}^{(i)}$(Line 11-14 in Algorithm~\ref{alg:k-hop}).
Through the exchange of edge information, the local model learns new information, 
and thus improves the accuracy of the global model. 

However, the PSI only preserves the privacy of the edges the other clients do not own. 
It's not able to protect the privacy of the edges in the intersection set. 
Thus, we leverage DP to \fedrec\ to extend \fedrec\ to \sys\ via DP.

\subsection{Privacy-Preserve User-Item Graph Sharing}

Since the user-item graph contains sensitive information involving user privacy, the direct interaction of the user-item graph between clients will be strictly restricted due to privacy regulations. 
\sys\ applies different DP algorithms in both topology as well as the weights information to preserve the privacy of both.

For the topology of the graph, \sys\ adds noises to a weighted undirected graph using the LAPGRAPH algorithm. 
For simplicity, we denote the user-item connection matrix of the graph as $M$, where 0 indicates that there is no scoring relationship between the corresponding user and the corresponding item, and the vice versa is 1. 
\sys\ first calculates the sparsity degree $T=n_1$ where $n_1$ is the number of 1.
Next, \sys\ adds Laplacian noise with a mean value of 0 and an intensity of $\lambda_1$ to each matrix element, and at the same time uses a part of the privacy budget (usually 1\%) to protect the sparsity degree $T$ from noise. 
We denote the sparse degree after adding noises is $T'$. 
Finally, \sys\ leaves the top $T'$ large elements of the matrix after adding noise as 1, and others as 0. 

For the protection of the edge weights information of the user-item graph, a Laplace noise with a mean value of 0 and an intensity of $\lambda_2$ is added directly to the new graph formed by the above algorithms.



\begin{algorithm}
\caption{K-hop extension of client $i$}
\label{alg:k-hop}
\begin{algorithmic}[1]
\Require $K$, the parameter of K-hop; the graph $G^i$
\Ensure the extended graph $\bar{G}^i$
\Procedure {K-hop Extension}{$K, i$}
\State $\bar{V}^i=V^i$, $\bar{E}^i=E^i$
\For{$u_j\in \mathcal{U}\backslash\{u_i\}$}
    \State $V^{i,j} = \mathsf{PSI}(V^i, V^j)$
    \State $\bar{V}^{i,j} = \bar{V}^{i,j} \cup \{v|\mathsf{dist}(v, V^{i,j}, G^i)\le k \land v \in V^i\}$
    \State $\bar{E}^{i,j} = \bar{E}^{i,j} \cup \{<x, y>| x,y \in \bar{V}^{(i,j)}\}$
    \State Send $(\bar{V}^{i,j},\bar{E}^{i,j}$) to client $j$
\EndFor

\For{$u_j\in \mathcal{U}\backslash \{u_i\}$}
    \State Receive ($\bar{V}^{j,i}, \bar{E}^{j,i}$) from client $j$
    \State $\bar{V}^i = \bar{V}^i \cup \bar{V}^{j,i}$
    \State $\bar{E}^i = \bar{E}^i \cup \bar{E}^{j,i}$
\EndFor

\State \textbf{return} $\bar{G}^i=(\bar{V}^{i}, \bar{E}^{i})$
\EndProcedure
\end{algorithmic}
\end{algorithm}

%% file: analysis.tex
\section{Analysis}

\subsection{Privacy Analysis}

The privacy of \sys\ is protected by the following aspects: 
(i) The vertexes outside of the intersection set during K-hop extension. The privacy of this part is guaranteed by PSI.
(ii) The vertexes within the intersection set during K-hop extension. The privacy of this part is guaranteed by DP. The protection of privacy is divided into protection of the topology structure and protection of the weights of the edges. For preservation of topology, the Laplace noise is added to its adjacency matrix using the Lapgraph algorithm so that the information is perturbed. For  protection of edge weights, the values of edge weights are disturbed by adding noises directly to the edge weights. It has been demonstrated in~\cite{wu2021linkteller} that when the noise  added satisfies $Lap(0, \frac{\Delta f_1}{\epsilon_1})$, the Lapgraph algorithm satisfies $\epsilon_1-DP$. At the same time, due to the characteristics of Laplace mechanism~\cite{Dwork2013privacybook}, the noise added to the edge weights satisfies $Lap(0,\frac{\Delta f_2}{\epsilon_2})$, which is $\epsilon_2-DP$. By Composition theorem~\cite{Dwork2013privacybook}, \sys\ satisfies $\epsilon_1+\epsilon_2$-DP.


\subsection{Performance Analysis}




First, consider the time complexity of \sys\ for one client. Since a certain amount of noise needs to be added to each element of the adjacency matrix, the time for single addition of noise is $\mathcal{O}(|V^i|^2)$. Then analysis is performed on the communication complexity of \sys\ for client $i$.
Since \sys\ requires interaction between two clients, communication cost of such interaction between clients is $\mathcal{O}(l^2)$. 
Each interaction contains PSI, K-hop extension, and adding noise towards expanded graph data. 
Correspondingly, the time complexity of PSI is $\mathcal{O}(|V^i|)$, the time complexity of K-hop extension is $O(|V^i|)$.
The time for single addition of noise, as analyzed above, is $\mathcal{O}(|V^i|^2)$. 
So the communication cost of \sys\ is $\mathcal{O}(l^2 \cdot |V^i|^2 )$.

%% file: evaluation.tex
\section{Evaluation}
\subsection{Evaluation Setup}
\subsubsection{Implementation}
We implement both \fedrec\ and \sys\ via Python based code of $\mathsf{\textsf{FedGraphNN}}$~\cite{he2021fedgraphnn}.
We conduct the evaluation on a computation instance equipped with 2.1GHz 64 Intel(R) Xeon(R Gold 6130 CPU, 512 GB memroy and 8 Tesla V100 GPU with 12GB.

\vspace{-5mm}

\subsubsection{Dataset}
We conduct evaluation on two datasets: Epinions~\cite{richardson2003trust} and MovieLens~\cite{harper2015movielens}. 
The Epinions dataset contains consumers’ ratings on items from the Epinions website.
The MovieLens dataset contains the users' rating of different movies from the MovieLens website. 
For both datasets, we divide the graph to different clients via the item category, \ie, the items with same category and their relevant users will be assigned to the same client. 
In particular, due to the large number of points in the epinions dataset and the limitation of memory, we only select 12 categories for experiments. Table~\ref{tab:dataset} shows the average number of users, items and edges after separation.

\vspace{-5mm}

\begin{table}[htbp]
\centering
\begin{tabular}{|c|c|c|c|c|c|}
\hline
Dataset & number of clients & K & \begin{tabular}[c]{@{}c@{}}Average n \\ for each client\end{tabular} & \begin{tabular}[c]{@{}c@{}}Average n\\ for each client\end{tabular} & \begin{tabular}[c]{@{}c@{}}Average \# edges\\ for each client\end{tabular} \\ \hline
\multirow{5}{*}{Epinions} & Centralized & / & 21296 & 163874 & 870838 \\ \cline{2-6} 
 & \multirow{2}{*}{8} & 2 & 21052 & 117641 & 754303 \\ \cline{3-6} 
 &  & 10 & 21172 & 163588 & 870266 \\ \cline{2-6} 
 & \multirow{2}{*}{12} & 2 & 21007 & 105897 & 720281 \\ \cline{3-6} 
 &  & 10 & 21165 & 163539 & 870227 \\ \hline
\multirow{5}{*}{MovieLens 1M} & Centralized & / & 6040 & 3706 & 1000209 \\ \cline{2-6} 
 & \multirow{2}{*}{8} & 1 & 5894 & 3704 & 995298 \\ \cline{3-6} 
 &  & 5 & 6040 & 3706 & 1000209 \\ \cline{2-6} 
 & \multirow{2}{*}{12} & 1 & 5409 & 3699 & 972759 \\ \cline{3-6} 
 &  & 5 & 6040 & 3706 & 1000209 \\ \hline
\end{tabular}
\caption{Dataset description. The $K$ is the parameter of K-hop, $n$ is the number of users, the $m$ is the number of items and \# edges is the number of edges. For centralized, the number of user, items and edges is the total number.}
\label{tab:dataset}
\vspace{-15mm}
\end{table}

\subsubsection{Setting of Experiments}
Our evaluation goal is to prove two claim: the K-hop extension improves the accuracy of the federated GNN and the leverage of DP in \sys\ do not reduce the accuracy too much.
The experiments is conducts under two client settings: 8 and 12 clients. Five experiments were performed in each setting: 
(i) Centralized training, the central server owns the full graph for training; 
(ii) \textsf{FedGraphNN} with FedAvg, the structure proposed in ~\cite{he2021fedgraphnn}, which is also the baseline we compared. For simplicity we denote it as \textsf{FedGraphNN} in the remaining sections;
(iii) \fedrec, where we only perform K-hop extension without adding noise to the interactive content. The purpose of this experiment is to demonstrate that the K-hop extension helps to increase the accuracy of link prediction;
(iv) \textsf{DP-FedGraphNN}, we add Laplace(0, 1) noise to the edge weights of the user-item graph based on \textsf{FedGraphNN} as a baseline to compare with \sys.
(v) \sys\ with different $K$, which is realized by adding noise to the interactive content on the basis of the third group of \fedrec.
For evaluation metrics, we adopt mean absolute error (MAE), mean square error (MSE) and root mean square error (RMSE) to evaluate the accuracy of edge weights prediction and record the average time it takes to add noise to a single client in each experiment.

\vspace{-5mm}
\begin{table}[htbp]
\centering
\begin{tabular}{|c|c|c|c|c|c|c|}
\hline
Dataset/8 clients & \begin{tabular}[c]{@{}c@{}}Model\\ Type\end{tabular} & System & MAE & MSE & RMSE & \begin{tabular}[c]{@{}c@{}}Noising\\ Time (s)\end{tabular} \\ \hline
\multirow{6}{*}{Epinions} & \multirow{3}{*}{W/O DP} & Centralized & 0.8377 & 1.2464 & 1.1164 & \multirow{3}{*}{/} \\ \cline{3-6}
 &  & \textsf{FedGraphNN} & 0.8719 & 1.3424 & 1.1559 &  \\ \cline{3-6}
 &  & \fedrec & 0.8643 & 1.3303 & 1.1505 &  \\ \cline{2-7} 
 & \multirow{3}{*}{W/ DP} & \textsf{DP-FedGraphNN} & 0.8724 & 1.3484 & 1.1584 & / \\ \cline{3-7} 
 &  & \sys(K=2) & 0.8689 & 1.3415 & 1.1560 & 328 \\ \cline{3-7} 
 &  & \sys(K=10) & 0.8658 & 1.3278 & 1.1523 & 517 \\ \hline
\multirow{6}{*}{MovieLens1M} & \multirow{3}{*}{W/O DP} & Centralized & 0.8812 & 1.1782 & 1.0855 & \multirow{3}{*}{/} \\ \cline{3-6}
 &  & \textsf{FedGraphNN} & 0.8832 & 1.1850 & 1.0884 &  \\ \cline{3-6}
 &  & \fedrec & 0.8793 & 1.1786 & 1.0884 &  \\ \cline{2-7} 
 & \multirow{3}{*}{W/ DP} & \textsf{DP-FedGraphNN} & 0.8912 & 1.2057 & 1.0980 & / \\ \cline{3-7} 
 &  & \sys(K=1) & 0.8820 & 1.1798 & 1.0862 & 4 \\ \cline{3-7} 
 &  & \sys(K=5) & 0.8813 & 1.1783 & 1.0875 & 4 \\ \hline
\end{tabular}
\caption{Performance of different systems with 8 clients. Noising time refers to the time of adding noise per client.}
\label{tab:performance_c8}
\vspace{-10mm}
\end{table}

\subsection{Performance of K-Hop Extension}

Table~\ref{tab:performance_c8} and Table~\ref{tab:performance_c12} show the performance of centralized server, \fedrec\ and  \textsf{FedGraphNN}.
It indicates that \fedrec\ performed better than \textsf{FedGraphNN} in all metrics. The result proves the K-hop extension does really help to handle the non-IID problem in federated learning and thus improve the link prediction accuracy. 

The effect of \sys\ is also better than \textsf{FedGraphNN}, which proves that the K-hop extension algorithm based on local differential privacy improves the performance of the model while protecting privacy. Meanwhile The K-Hop extension is very robust even with adding noise to the graph data.

\subsection{Performance of Differential Privacy}

From the results, the performance of \sys\ does not decrease much than \fedrec. However, after adding noise to \textsf{FedGraphNN}, accuracy drops a lot.

\vspace{-5mm}

\begin{table}[htbp]
\centering
\begin{tabular}{|c|c|c|c|c|c|c|}
\hline
Dataset/12 clients & \begin{tabular}[c]{@{}c@{}}Model\\ Type\end{tabular} & System & MAE & MSE & RMSE &  \begin{tabular}[c]{@{}c@{}}Noising\\ Time (s)\end{tabular} \\ \hline
\multirow{6}{*}{Epinions} & \multirow{3}{*}{W/O DP} & Centralized & 0.8377 & 1.2464 & 1.1164 & \multirow{3}{*}{/} \\ \cline{3-6}
 &  & \textsf{FedGraphNN} & 0.8674 & 1.3279 & 1.1502 &  \\ \cline{3-6}
 &  & \fedrec($K=10$) & 0.8635 & 1.3270 & 1.1496 &  \\ \cline{2-7} 
 & \multirow{3}{*}{W/ DP} & \textsf{DP-FedGraphNN} & 0.8716 & 1.3298 & 1.1513 & / \\ \cline{3-7} 
 &  & \sys($K=2$) & 0.8625 & 1.3261 & 1.1493 & 314 \\ \cline{3-7} 
 &  & \sys($K=10$) & 0.8585 & 1.3258 & 1.1493 & 501 \\ \hline
\multirow{6}{*}{MovieLens1M} & \multirow{3}{*}{W/O DP} & Centralized & 0.8812 & 1.1782 & 1.0855 & \multirow{3}{*}{/} \\ \cline{3-6}
 &  & \textsf{FedGraphNN} & 0.8931 & 1.2454 & 1.1152 &  \\ \cline{3-6}
 &  & \fedrec($K=5$) & 0.8874 & 1.1850 & 1.0883 &  \\ \cline{2-7} 
 & \multirow{3}{*}{W/ DP} & \textsf{DP-FedGraphNN} & 0.8989 & 1.2669 & 1.1247 & / \\ \cline{3-7} 
 &  & \sys($K=1$) & 0.8936 & 1.2257 & 1.1063 & 3 \\ \cline{3-7} 
 &  & \sys($K=5$) & 0.8907 & 1.1991 & 1.0948 & 4 \\ \hline
\end{tabular}
\caption{Performance of different systems with 12 clients.  Noising time refers to the time of adding noise per client.}
\label{tab:performance_c12}
\vspace{-10mm}
\end{table}

When the number of clients in the Epinions dataset is 12, the effect of \sys\ is better than that of \fedrec, which to a certain extent shows that the noise added in the experiment not only protects the privacy of the data from being leaked, but also ensures the data availability is not compromised, reflecting the balance between data privacy and availability.

We also recorded the average time for each client to add noise.
According to our analysis, the time for adding noise is positively correlated with the number of points in the graph, \ie, the number of points increases, the time it takes to add noise will increase, while the increase in the number of edges does not significantly increases the time it takes to add noise. 

In the Epinions dataset and MovieLens1M dataset, the number of points of Epinions is significantly larger than that of MovieLens1M, and the number of edges of MovieLens1M is significantly larger than the number of points of Epinions. Summarizing the average time to add noise per client in the experiments, we found that the time required to add noise to the Epinions dataset is much greater than that required for Movielens which is consistent with our analysis. 

%% file: related_work.tex
\vspace{-2mm}
\section{Related Work}
\subsection{Federated Recommendation System}

Federated Learning is being applied in lots of field~\cite{2021Modeling,2020Network,si2022federated,sun2022fair,li2022federated}. And as the laws and regulations of data and privacy become stricter, recommendation systems based on federated learning with privacy-preserving features have become a hot research trend. 
FCF~\cite{ammad2019federated}, a classic federated recommendation system, is the first collaborative filtering framework based on the federated learning paradigm. They build a joint model by using user implicit feedback. 
\cite{qi2020privacy} is a privacy-preserving method which leverages the behavior data of massive users and meanwhile don't require centralized storage to protect user privacy to train news recommendation model with accuracy.
FedFast~\cite{2020FedFast} achieves high accuracy for each user during the federated learning training phase as quickly as possible. In each training round, They sample from a set of participating clients and apply an active aggregation method that propagates the updated model to the other clients.

\subsection{Differential Privacy Graph Neural Network}
Several literature leverage DP to preserve the privacy of GNN. 
Solitude~\cite{LDP2022Wanyu} is a privacy-preserving learning framework based on GNN, with formal privacy guarantees based on edge local differential privacy. The crux of Solitude is a set of new delicate mechanisms that calibrate the introduced noise in the decentralized graph collected from the users. 
LDPGen~\cite{Qin2017LDP} is a multi-phase technique that incrementally clusters users based on their connections to different partitions of the whole population. LDPGen carefully injects noise to ensure local differential privacy whenever a user reports information.
There are only few works that combine the DP with the GNN federated learning.
\cite{wu2021fedgnn} applies differential privacy techniques to the local gradients of GNN model to protect user privacy in federated learning setting. But it need a third-party server to store embedding of users besides training server.So FedGNN is a two-server model.
In \cite{zhou2020vertically}, They propose (VFGNN), a federated GNN learning model for privacy-preserving node classification task under data vertically partitioned setting. They leave the private data related computations on data holders, and delegate the rest of computations to a semi-honest server. However, their work has an strong assumption that every data holders have the same nodes, which is far different from real scenario.

%% file: conclusion.tex
\section{Conclusion and Future Work}
In this paper, we proposed \sys\ a privacy-preserving federated GNN framework for recommendation system.
To overcome the challenge of the Non-IID problem under the privacy regulation, \sys\ integrates the PSI and the DP technique with the federated GNN. 
The PSI-based K-hop extension helps to extend the sub-graph of each client without leaking any non-intersection information to solve the non-IID problem.
Moreover, DP preserves not only the privacy of weights but also the privacy of edges/topology in the intersection information to guarantee the privacy.
We implemented the prototype of \sys\ and tests it on different datasets. 
Compared with other works, the evaluation shows \sys\ achieves high performance and only induces little computations overhead.
In the future, we would like to investigate a universal DP for both weights and edges in graph data for better performance.

\section{Acknowledgement}\label{Acknowledgement}
This paper is supported by the Key Research and Development Program of Guangdong Province under grant No.
2021B0101400003. Corresponding author is Jianzong Wang from Ping An Technology (Shenzhen) Co., Ltd (jzwang@188.com).



